\documentclass[a4paper,fleqn]{cas-dc}

\usepackage[authoryear,longnamesfirst]{natbib}

\def\tsc#1{\csdef{#1}{\textsc{\lowercase{#1}}\xspace}}
\tsc{WGM}
\tsc{QE}

\begin{document}
\let\WriteBookmarks\relax
\def\floatpagepagefraction{1}
\def\textpagefraction{.001}

\shorttitle{}    

\shortauthors{}  

\title [mode = title]{CALM: Cognitive Assessment using Light-insensitive Model}  



%

\author[1]{Akhil Meethal}

\cormark[1]


\ead{akhilpm135@gmail.com}


\author[1,2]{Anita Paas}
\author[1]{Nerea Urrestilla Anguiozar}
\author[1]{David St-Onge}

\affiliation[1]{organization={ETS Montreal},
            addressline={ University of Quebec}, 
            country={Canada}}






\affiliation[2]{organization={Department of Psychology, Concordia University},
country={Canada}}

\cortext[1]{Corresponding author}




\begin{abstract}
The demand for cognitive load assessment with low-cost easy-to-use equipment is increasing, with applications ranging from safety-critical industries to entertainment. Though pupillometry is an attractive solution for cognitive load estimation in such applications, its sensitivity to light makes it less robust under varying lighting conditions. Multimodal data acquisition provides a viable alternative, where pupillometry is combined with electrocardiography (ECG) or electroencephalography (EEG). In this work, we study the sensitivity of pupillometry-based cognitive load estimation to light. By collecting heart rate variability (HRV) data during the same experimental sessions, we analyze how the multimodal data reduces this sensitivity and increases robustness to light conditions. In addition to this, we compared the performance in multimodal settings using the HRV data obtained from low-cost fitness-grade equipment to that from clinical-grade equipment by synchronously collecting data from both devices for all task conditions. Our results indicate that multimodal data improves the robustness of cognitive load estimation under changes in light conditions and improves the accuracy by more than 20\% points over assessment based on pupillometry alone. In addition to that, the fitness grade device is observed to be a potential alternative to the clinical grade one, even in controlled laboratory settings.
\end{abstract}
\nocite{*}


\maketitle

\section{Introduction}
\label{sec:intro}
The level of automation in task performance is increasing across various domains, including manufacturing, transportation, and healthcare \cite{automation-Coombs-2019}. Concurrently, human involvement has shifted towards the verification of activities rather than direct task completion \cite{work_design-Parker-2022}. In most applications, humans still bear the pressure of success or failure, even if they are not directly controlling the automation. They must verify and ensure the correct functioning of automated tasks indirectly. Given this trend, understanding Cognitive Load (CL) during task performance is receiving increased attention. This must be performed in real-time in uncontrolled environments where the human operator is performing the task. Thus, reliable Cognitive Load Estimation (CLE) using devices that won't obstruct the task performance is paramount.   

The increasing availability of low-cost, non-invasive devices makes it easier to measure cognitive response in real time during activities. Currently, several commercial devices measure pupillometry, heart rate variability (HRV), skin conductance, electroencephalography, and body motion \cite{mobile_measure-Caldiroli-2023, low_cost_cogload-Rajat-2014}. Among these, pupillometry provides an attractive option due to its low cost, unobstructive during the task and ease of setup and use, especially in field applications. But, Pupillary Response (PR) can be influenced by changes in light conditions, which should not be misattributed to cognitive load \cite{pupilary_response-Arthur-1991}. This is particularly important when measuring cognitive load in uncontrolled environments where light conditions can vary significantly. Research indicates that combining pupillometry with other physiological and behavioral measures offers a more comprehensive assessment of cognitive load \cite{multimodal_cogload-Vanneste-2020}.  

While the existing studies employed multimodal data for cognitive load estimation, focused studies on understanding the reliability and robustness of the pupillometry-based multimodal CLE under varying light conditions are lacking. Also, the studies in controlled environments (simulation-based laboratory studies) mainly used clinical-grade devices \cite{multimodal_cle_biopac-Oppelt-2023} whereas the uncontrolled field studies used fitness-grade devices \cite{multimodal_filed-David-2024} for the HRV. A comparison of both types of devices in the same settings to understand their characteristics is also absent in the current literature. The primary objective of this research is to understand the impact of light conditions on pupillometry-based CLE and quantify how a multimodal approach, combined with HRV, is improving the reliability and robustness. We used HRV because it is invariant to light conditions. In addition to the primary objective, our study includes a secondary objective aimed at comparing the low-cost, easy-to-use fitness-grade devices for HRV data in multimodal settings over clinical-grade HRV devices. 


\subsection{Pupillometry and the light condition}
The variation of the pupil diameter when performing a task is widely used as a cognitive load estimator. Comparison of the cognitive load estimation results from gaze tracking and pupillary response identified the benefits of both eye movement-based measures \cite{pd_gaze_cogload-Krejtz-2018}. Even though it is among the most recent quantitative cognitive metrics, pupillometry accounts for a significant amount of research \cite{pupil_cogload-Weber-2021, PupillometryPP-Matht-2018}. The utility of pupillometry as a mental load estimator stems from the pupil's sensitivity to physiological factors related to the autonomic nervous system (ANS). Pupillometry is non-invasive and barely affects the task performance, making it an attractive choice for field applications. 

One of the main drawbacks of pupillometry is that pupil diameter changes according to external light conditions. This means that features derived from pupil diameter are influenced not only by cognitive load but also by ambient light. To mitigate this sensitivity, Duchowski et al. developed the Index of Pupillary Activity (IPA) and the Low/High Index of Pupillary Activity (LHIPA) features \cite{Duchowski2018, Duchowski2020}. The Index of Cognitive Activity (ICA) is also widely recognized as less sensitive to light \cite{ica-Marshall-2002}. However, many recent studies have observed that these features still suffer from variations in light conditions \cite{light_compare-Nickolas-2024}. Our experiments also confirmed the sensitivity of pupillometry-derived features to ambient light, leading us to adopt a multimodal approach to reduce this sensitivity. 

\subsection{Multimodal CLE with Pupillometry and HRV}
HRV is a well-studied metric in cognitive load estimation \cite{Solhjoo2019}. HRV is relatively robust to most external influences and is easy to measure and post-process. This measure is favored because of its direct relation to the ANS. The interaction between the sympathetic and parasympathetic nervous systems, the two main components of the ANS, is identified in heart activity as low-frequency (LF) and high-frequency (HF) bands. The balance between these two frequencies, known as the sympathovagal balance, is directly related to cognitive load \cite{hrv_sympathovogal-Kathrin-2021}.

HRV is sensitive to variations in the electromagnetic field and breathing cycle \cite{Shaffer2017}. HRV signal acquisition can also be affected when the operator is moving, due to touch intensity variations leading to fluctuations in the recorded signal. Using a multimodal approach, we seek to extract complementary information from electrocardiography (ECG) and pupillometry for robust cognitive load estimation. 


To evaluate the feasibility of using fitness grade HRV devices over clinical grade ones, we chose to collect HRV from two distinct devices at the same time: the Biopac MP35 and the Polar H10. The Biopac MP35 is a sophisticated HRV data-collection tool predominantly used in clinical research contexts. This device requires participants to remain immobile throughout data collection, thereby restricting participant movement. Conversely, the Polar H10 serves as a heart rate monitoring device designed specifically for use during physical activity and exercise \cite{wearable_ECG-Alugubelli-2022}. The Polar H10 communicates wirelessly with the data acquisition system, facilitating seamless data collection even during activities. To compare the Polar and Biopac devices fairly under the same experimental conditions, we combined the pupillometry data with HRV data from each device separately and assessed the cognitive load estimation performance. The research objectives of this study can be summarized as follows:

\begin{itemize}
    \item How does the change in light conditions affect the performance of pupillometry-based CLE? Combined with HRV, how can the multimodal data improves the reliability and robustness against change in light conditions? 
    \item Is there a significant difference in CLE performance when low-cost fitness-grade devices are used for HRV data acquisition over clinical-grade ones in a multimodal setting?
\end{itemize}


\section{Methods and Materials}
\label{sec:method}
To validate our objectives, we collected participants' HRV and pupillometry data under light and dark conditions. Participants were asked to perform the n-back task to vary the level of cognitive load \cite{n-back_main-Kirchner-1958, n-back_modern-Kane-2007}. We conducted extensive experiments varying the cognitive load and light conditions. Multimodal data is acquired from all settings using both HRV devices. 

\subsection{Participants} 
\noindent We recruited 10 participants (5 female, 5 male) from the research group. All participants were between 20 to 45 years old. This research was approved by the ÉTS (Ecole de Technology Supérieure) Research Ethics Board in accordance with the Canadian Tri-Council Policy Statement regarding the Ethical Conduct for Research Involving Humans (TCPS2; Approval number H20210302). 

\subsection{Experimental Design}

\subsubsection{Sensors}
\begin{figure}[!h]
  \centering
  \includegraphics[width=0.6\columnwidth]{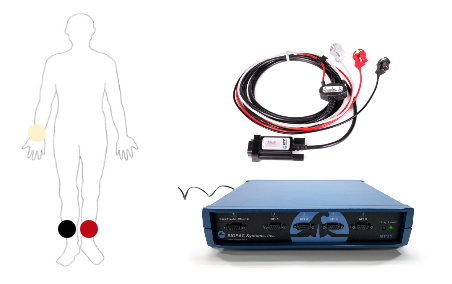}
  \includegraphics[width=0.35\columnwidth]{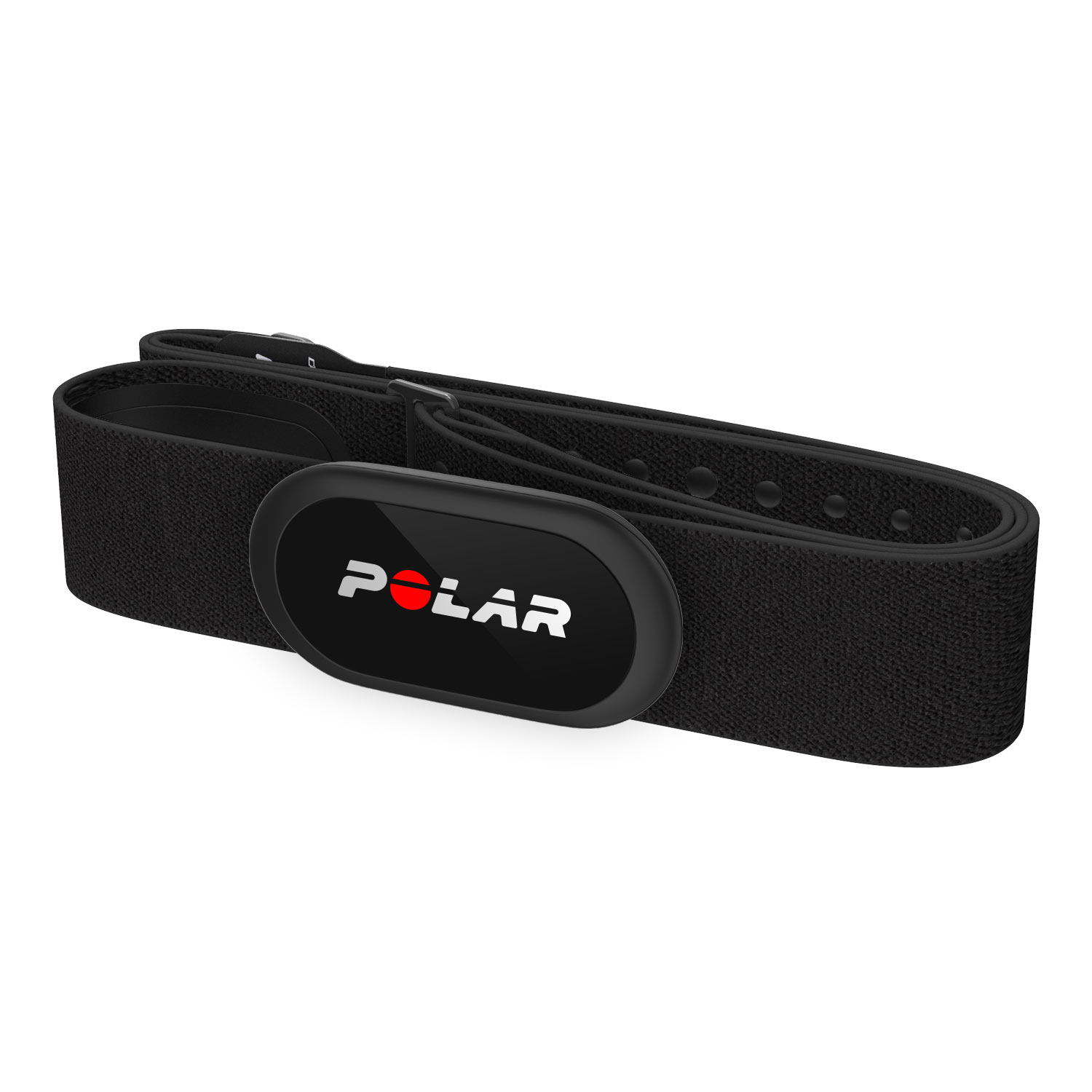}
  \caption{HRV devices: Left, BioPac M35 with the electrodes position, Right, Polar H10 chest band used only in the second experiment.}
  \label{fig:hrvdev}
\end{figure}

\begin{figure}[!h]
  \centering
  \includegraphics[width=0.57\columnwidth]{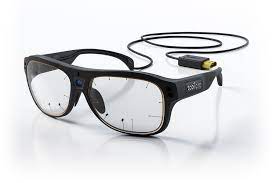}
  \caption{Pupillometry device: Tobii Pro 3 glasses}
  \label{fig:pupildev}
\end{figure}

\noindent The sensors used for recording the data were Tobii glasses along with two ECG devices: the BioPac MP35 and the Polar H10 chest band. The Polar H10 (see Fig. \ref{fig:hrvdev}; Polar, Electro, Oy, Finland) is a wireless chest band equipped with two electrodes positioned one inch apart on the chest to measure ECG. Although it is expected to be less accurate compared to clinical-grade devices, it offers the advantages of portability and ease of use. The BioPac MP35, on the other hand, is a three-electrode device that provides high-fidelity data collection for measuring HRV. We used a single analog channel with three electrodes placed on the body as illustrated in Fig. \ref{fig:pupildev}: right arm (white), right leg (black), and left calf (red). This configuration allows for triangulation to extract a clean ECG signal. The main drawback of the BioPac MP35 is that it is wired, which limits its field of application, but it provides a more detailed dataset compared to the Polar H10.

For measuring and recording pupillometry, Tobii Pro 3 glasses were used (see Fig. \ref{fig:pupildev}). It is a wireless eye-tracking platform that contains cameras pointing at the pupil mounted on a glass frame with illuminators. It is able to record and perform the calibration automatically. The data was streamed over WiFi to the researcher's machine, on which all the data was stored and synchronized. The glasses are wired to the streaming and battery-containing device which is attached to the waist of the participant.

\subsubsection{Procedure} 
\noindent Experiments took place in a laboratory environment, where the lighting conditions could be controlled. Participants came to the testing room and the task and equipment were explained to them. After consenting to continue with the experiment, participants were instructed how and where to place the Polar H10 and then shown to the bathroom where they could dampen the electrodes and place the chest band underneath their shirt across their chest with the electrodes at their sternum. Next, the BioPac MP35 was put on. Three 3M Red Dot electrodes with a built-in abrader (3M, ID: 7000128699) were placed on the recently cleaned right wrist and both ankles. Finally, the Tobii Pro 3 glasses were put on. The glasses were then calibrated according to the procedures provided by Tobii to ensure accurate measuring and recording of eye data. After all the devices were set up, the experimenter started recording the data, synchronized across devices using ROS. Once the experimenter ensured that all the devices were recorded properly, the n-back task (described below) began. Participants were comfortably seated in a chair throughout each experiment.

\begin{figure}[!h]
  \centering
  \includegraphics[scale=0.38]{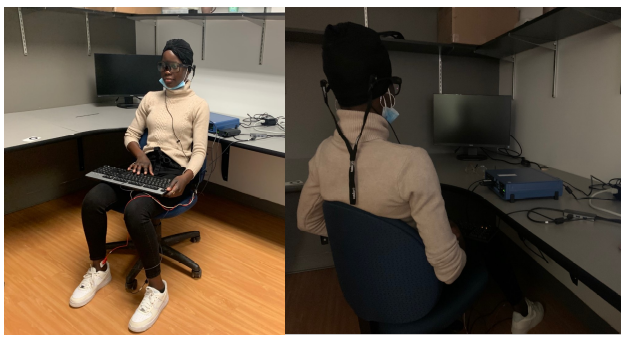}
  \caption{Left: Experiment setup for the Light condition. Right: Experiment setup for the Dark condition.}
  \label{fig:setup}
\end{figure}

There were three levels of mental workload (rest, low, and high) and two levels of lighting (light [210 Lux] and dark [1 lux]). In the rest condition, participants looked at a point on the wall and relaxed while their pupil and heart rate data were recorded. During the low workload condition, participants performed the 1-back task, at which they heard a sequence of stimuli (numbers between 0 and 26) and pressed a key on the keyboard whenever they heard a match between the current stimulus and the stimulus presented previously. In the high workload condition, participants performed a 2-back task \cite{n-back_main-Kirchner-1958}. In the 2-back task, participants heard a sequence of stimuli - numbers between 0 and 26. Each time participants heard a number that matched the number presented 2 stimuli previously in the sequence, they pressed a key on the keyboard to indicate they heard a match. This task induces a high mental workload because it requires participants to maintain and continuously update two digits in their working memory. The duration of the task was 3 minutes. Between each condition, there was a break of at least 2 minutes (longer if the participant needed more time). The order of conditions (both workload and lighting) was counterbalanced across participants. Figure \ref{fig:setup} shows the setup for this experiment, with the light condition on the left and the dark condition on the right. The whole experiment duration was approximately one hour. 

\subsection{Data Pre-processing and Feature Extraction}
\subsubsection{Pupillometry}

Pupil diameter data are usually noisy due to eye movements and blinking, which results in missing data blocks. Packet loss during recording also results in missing data. The missing data were handled by re-sampling the signal, keeping a fixed sampling frequency (set to 100Hz). As a first step towards cleaning the data, outliers were removed, keeping the diameter range between 1.5 and 9 mm (the general minimum and maximum for pupil diameter). The noise was handled using filtering and smoothing. A low-pass butter filter of order 5 and a cutoff of 4 Hz was used in this process \cite{butter_filter-Knapen-2016}. We used a sliding window of 60 seconds and the pupil diameter values were averaged over each window. 

Within each window, the recorded pupil diameter values were used to compute the required features. After careful consideration of the significance of the features in predicting cognitive load, the following features were selected: pupil diameter, standard deviation of the pupil, Index of Pupillary Activity (IPA), PDAuC (Pupil Diameter Area under the Curve), and PDRoC (Pupil Diameter Rate of Change). PDRoC finds the slope of the regression line fit on pupil size data. Both PDAuC and PDRoC were obtained using the PyTrack library \cite{pytrack-Tortora-2022}. IPA is computed following the implementation from \cite{ipa-Duchowski-2018}

\subsubsection{Heart Rate Variability}
Heart-related data were recorded by two different devices during the experiments - the BioPac MP35 and the Polar H10. The HRV features were extracted from the raw signals of both devices, over a window of 60 seconds. Sliding windows with an overlap of 10 seconds between windows were used for feature extraction. For Polar H10, the sampling rate was 120 Hz and the signal was resampled to account for the loss of data. For BioPac MP35, the sampling rate was 1000 Hz. The signal values outside the range [300-1000] mV were considered outliers and they were removed. Finally, resampling was performed to make the session length consistent to a fixed 3-minute duration.

The feature extraction was carried out using the neurokit2 toolbox \cite{neurokit2-Makowski-2021}. For this, RR peaks (the time between successive R waves) were first computed from the raw ECG data after pre-processing. Once this was completed, the next step was to compute the time domain and the frequency domain features. From the time domain, the features extracted were RMSSD (root mean square of successive differences), SDNN (standard deviation of normal-to-normal intervals), pNN50 (percentage of successive normal RR intervals that differ by more than 50ms), mean RR interval, median RR interval, and respiration rate. From the frequency domain, the features extracted were HF (high frequency) and LFHF Ratio (low-frequency power to high-frequency power). 

\subsection{Classifiers}\label{met:class}
The classifier used in our experiments was primarily the random forest (RF) classifier, but we also compared the results with neural network-based classifiers. The RF classifier was chosen due to its robustness, the feasibility of doing feature importance analysis, and low risk of over-fitting, especially with smaller datasets. In the RF classifier, 100 estimators were used. We used classification accuracy on the test set as the main metric for comparison as it provides a direct measure of how well the model generalizes to unseen data. High accuracy on a test set indicates good generalization assuming the test set is representative of the real-world data the model will encounter (commonly used assumption in machine learning classifiers). A 70-10-20 split was used for training, validation, and test set respectively. 

We selected two neural network-based classifiers for the comparative study, with and without using hand-crafted features. The first was a multi-layer network (called MLP or multi-layer perceptron) trained on the same features used with the RF classifier (with hand-crafted features). The second was a transformer-based architecture \cite{transformer-Vaswani-2017} which learns the features from the raw input signals (without hand-crafted features). The learnable features eliminate the need for deep domain expertise required to design and verify the hand-crafted features \cite{feat_learning-Bengio-2013}. The MLP is a 4-layer network with ReLU activation and batch normalization \cite{batch_norm-Sergey-2015}. The hidden layer dimensions are 256, 128, and 64, respectively, from layers 1 to 3. The Transformer classifier is a custom-designed architecture with 4-layer attention-based network \cite{transformer-Vaswani-2017} with GeLU activation and LayerNorm \cite{layernorm-Xu-S2019}. To manage the large sequence length at the input of the Transformer, a 7-layer 1D convolutional network is employed initially to obtain compact embeddings from the input, inspired from \cite{wave2vec-Baveski-2020}.

\section{Results}
\label{sec:experiments}
The extent of our experiments allows us to address several topics from the analysis of the data. We start by analyzing the overall performance difference due to change in light conditions and how the multimodal data is improving it. Then we go for a more fine-grained analysis at the feature level when light conditions vary. All models in this section are trained with an RF classifier using 100 estimators. 

\subsection{Pupillometry and change in light conditions}
To highlight the sensitivity of pupillometry to light, we trained classifiers under different lighting conditions and tested them separately in light and dark conditions. Specifically, five settings were considered for training and evaluation. The training set includes both light-only conditions (labeled as "Light") and mixed light and dark conditions (labeled as "All"). The test set includes light-only, dark-only (labeled as "Dark"), and mixed light and dark conditions. This split aims to understand the impact of light on pupillometry-based cognitive load estimation.

\begin{table}
    \caption{Comparison of the accuracy between Pupillometry only and multimodal (Pupillometry+HRV) data on all classifiers (RF, MLP and Transformers).}
    \centering\footnotesize
    \begin{tabular}{ l c c c c c}
    \toprule
    \textbf{Sensors} & \textbf{Train} & \textbf{Test} & \textbf{RF} & \textbf{MLP} & \textbf{Trans.} \\
    \midrule
    \midrule
    Pupil & Light & Light & 79.46 & 51.16 & 56.94  \\
    Pupil & Light & Dark & 46.56 & 34.38 & 37.50  \\
    Pupil & All & Light & 80.30 & 55.53 & 58.33 \\
    Pupil & All & Dark & 73.67 & 38.75 & 52.78 \\
    Pupil & All & All & 77.05 & 53.35 & 55.14 \\
    \midrule
    HRV(B)+Pupil & Light & Light & 92.24 & 82.81 & 86.11  \\
    HRV(B)+Pupil & Light & Dark & 47.71 & 38.88 & 41.67 \\
    HRV(B)+Pupil & All & Light & 91.53 & 81.25 & 84.72 \\
    HRV(B)+Pupil & All & Dark & 83.59 & 62.50 & 79.17 \\
    HRV(B)+Pupil & All & All & 89.42 & 78.95 & 84.66 \\
    \midrule
    HRV(P)+Pupil & Light & Light & 96.29 & 84.38 & 92.19 \\
    HRV(P)+Pupil & Light & Dark & 50.77 & 39.25 & 38.19 \\
    HRV(P)+Pupil & All & Light & 95.46 & 83.59 & 91.41 \\
    HRV(P)+Pupil & All & Dark & 91.98 & 65.63 & 79.16 \\
    HRV(P)+Pupil & All & All & 92.26 & 81.86 & 90.23 \\
    \bottomrule
    \end{tabular}
    \label{tab:three_load_results}
\end{table}

The results of the cognitive load classification task are reported in table \ref{tab:three_load_results}. In the table, HRV(B) and HRV(P) denote the HRV features computed from Biopac and Polar devices respectively. First, we will analyze the results of the RF classifier. In all cases, we see a significant drop in performance (33-45\% points) when trained on light conditions and tested on dark conditions, clearly showing how the distribution shift due to light is impacting the cognitive load estimation. The performance of the pupillometry only model decreases despite using features like IPA to reduce the sensitivity to light, as noted in \cite{light_compare-Nickolas-2024}. When the training set has data from both light and dark conditions, the accuracy drop is less (in the range of 4-9\% points). Training and testing in the light condition has the best performance in all cases. 

\subsection{Multimodal models}
The core strategy envisioned to reduce the lighting influence over pupillometry is to train on multiple modalities together. We thus used multimodal input to train a classifier combining HRV features with pupillometry features under the same conditions as previously described. The results are shown in the lower parts of Table \ref{tab:three_load_results}. It is evident that using multimodal inputs significantly enhances the classifier's performance. An overall improvement of 12.4\% and 15.2\% is observed with Biopac and Polar, respectively, using the multimodal features. 

Training in light and testing in dark conditions incurs significant performance loss. With pupillometry alone, this loss is around 33\% points. Although the multimodal settings show slight improvement compared to the pupillometry-only case,  the performance remains lower than the results when the training set contains data from all light conditions. This implies that the distribution shift due to light conditions still impacts performance as the pupillometry features show significant variance. Thus, we can conclude that even though multimodality improves the accuracy of CLE, the impact of distribution shift due to light cannot be neglected. Therefore, it is crucial to include data from diverse lighting conditions in the training set to achieve robust cognitive load estimation.

\begin{figure}[!ht]
\centering
\begin{tabular}{c @{\hspace{0.1cm}} c  @{\hspace{0.1cm}} c @{\hspace{0.1cm}} c}
\includegraphics[width=0.28\linewidth, height=2.4cm]{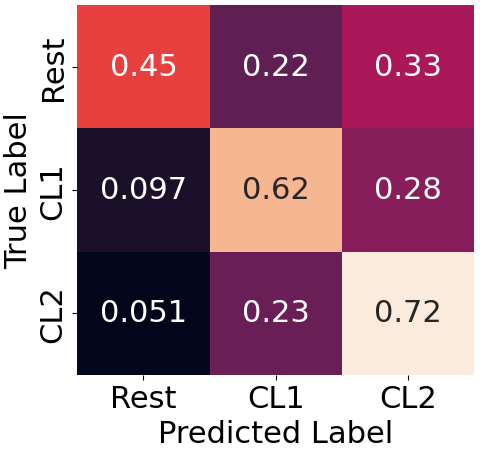} &
\includegraphics[width=0.28\linewidth, height=2.4cm]{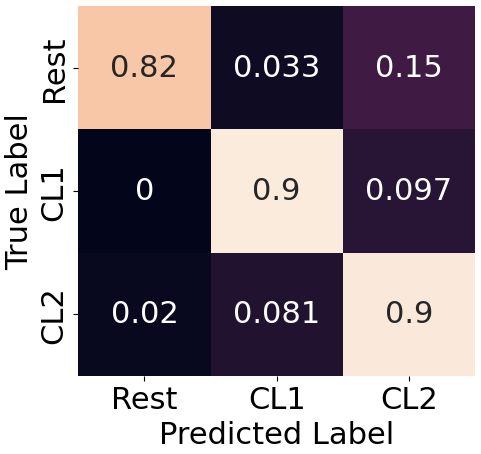} &
\includegraphics[width=0.28\linewidth, height=2.4cm]{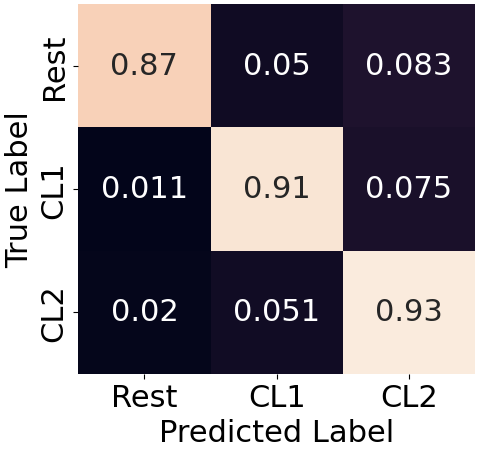} & 
\includegraphics[width=0.08\linewidth, height=2.4cm]{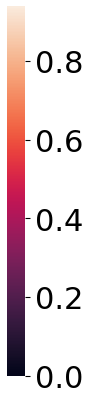} \\
Pup. & HRV(B)+Pup. & HRV(P)+Pup. & \\
 \end{tabular}
 \caption{Comparison of confusion matrices between pupillometry only and multimodal models.}
 \label{fig:cm_comparison}
 \vspace{-0.4cm}
\end{figure}

In figure \ref{fig:cm_comparison}, we compared the confusion matrices of the predictions from pupillometry and the two multimodal settings. When pupillometry alone is used, the classifier struggles to distinguish between the mental workload levels. Additionally, the Rest state shows significant confusion with other states. However, with multimodal data, the mental workload levels are more clearly separated by both Biopac and Polar sensors. Polar reduces the confusion of all states more than the Biopac.

\subsection{Impact of HRV device type}
When comparing the Biopac MP35 and the Polar H10 under our experimental conditions, the Polar H10 yielded the best results. The overall accuracy difference between the two was not substantial, around 2.8 percentage points. Although Biopac offers high-fidelity data collection suitable for clinical research applications, its wired setup is sensitive to participant movements, resulting in noisy data. In our experiments, participants were seated in a chair and exhibited movement while performing the n-back task, affecting the data quality. Conversely, Polar, designed for activity tracking, facilitated easier data collection. Based on the observations in Table \ref{tab:three_load_results}, we conclude that Polar is as good as Biopac or even better in experimental conditions where the participants are comfortably seated during data collection but may still exhibit movement. In adaptive automation contexts, workers are expected to be mobile, moving between stations and adjusting their working positions to prevent injuries or illnesses from prolonged sitting.

An overall comparison of both individual HRV device performances and their combination with pupillometry is reported in Table \ref{tab:overall_biopac_pupil}. The training and test sets included data from both light and dark conditions. The results show only a small difference of 1.2\% in cognitive load estimation accuracy between Biopac and Polar. When combined with pupillometry features, the accuracy improved by approximately 10\% points compared to using Biopac and Polar alone. The multimodal settings also showed increased performance compared to pupillometry-only results, with an accuracy improvement of over 12\% points for Biopac and 15\% for Polar. Additional material with more experiments using PupilLabs and Biopac sensors on a binary cognitive load classification problem (Rest vs Load) is provided here:  \href{https://initrobots.ca/calm/}{https://initrobots.ca/calm/}

\begin{table}
    \caption{Accuracy comparison of all models with full data under single-modal and multimodal settings.}
    \centering\footnotesize
    \begin{tabular}{ l c c c}
    \hline
    \textbf{Sensors} & \textbf{RF} & \textbf{MLP} & \textbf{Transformer} \\
    \hline
    \hline
    Pupil & 77.05$\pm$0.40 & 53.35$\pm$0.18  & 55.14$\pm$p0.34 \\
    HRV(B) & 80.96$\pm$0.41 & 74.61$\pm$0.21 & 79.04$\pm$p0.32 \\
    HRV(P) & 82.15$\pm$0.28 & 75.20$\pm$0.25 & 83.98$\pm$0.29 \\
    HRV(B)+Pupil & 89.42$\pm$0.28 & 78.95$\pm$0.22 & 85.66$\pm$0.38 \\
    HRV(P)+Pupil & 92.26$\pm$0.19 & 81.86$\pm$0.17 & 90.23$\pm$0.28\\
    \hline
    \end{tabular}
    \label{tab:overall_biopac_pupil}
\end{table}

\subsection{Comparison with Neural Net models}
Table \ref{tab:overall_biopac_pupil} also presents the comparison of neural network classifiers with the RF classifier. The advantages of multimodal signals over a single modality are evident from these results as well, wherein all cases, using multimodal signals yielded the best results. Although the light-insensitive HRV sensors are superior to pupillometry alone, pupillometry still provides valuable information that further enhances the CLE. This is due to the sensitivity of HRV measurements to participant movement. The Transformer-based classifier with learned features shows better results than the MLP in all cases. Except for pupillometry, the results of the Transformer are comparable to the RF classifier. It is widely known that neural networks are sensitive to distribution shifts \cite{domain_shift_prot-Meethal-2024, cross_domain_shift-Li-2022}. Given that we explicitly captured shifted data distributions for pupillometry, the Transformer models are expected to perform less effectively compared to RF. In table \ref{tab:three_load_results}, a detailed analysis of the performance of all classifiers under varying lighting conditions is provided.


\subsection{Feature-level Analysis}

\begin{figure*}[!ht]
\centering
\begin{tabular}{c @{\hspace{-0.3cm}} c  @{\hspace{0.2cm}} | c}
\includegraphics[width=0.33\linewidth, height=3.5cm]{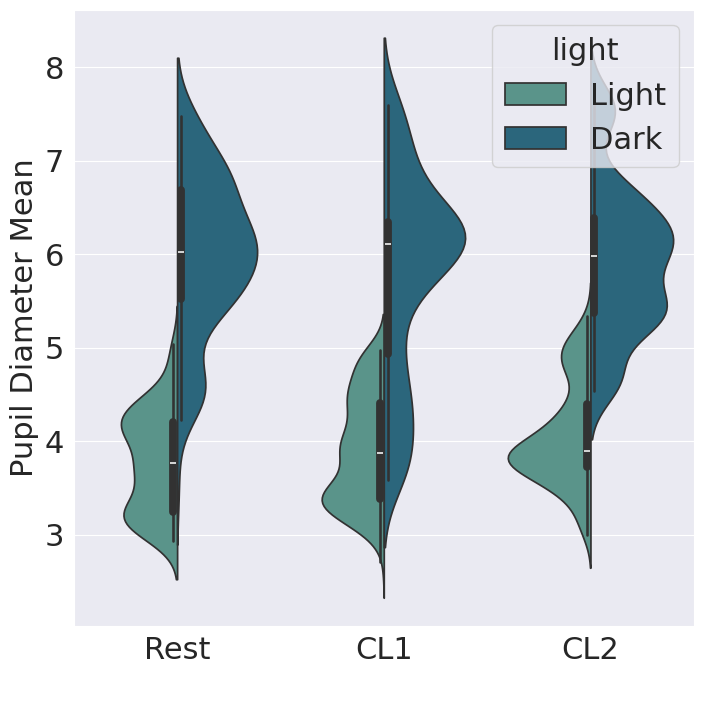} &
\includegraphics[width=0.33\linewidth, height=3.5cm]{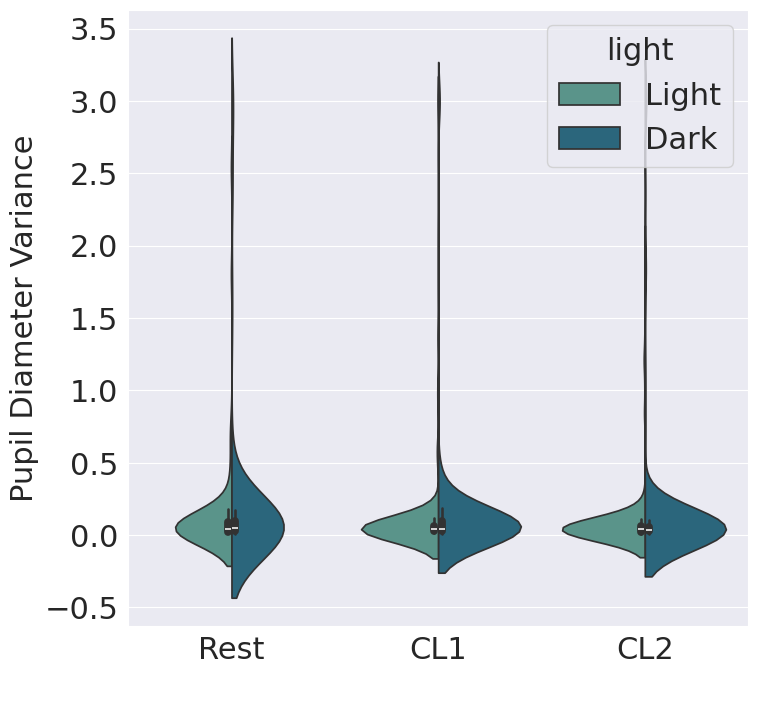} &
\includegraphics[width=0.32\linewidth, height=3.5cm]{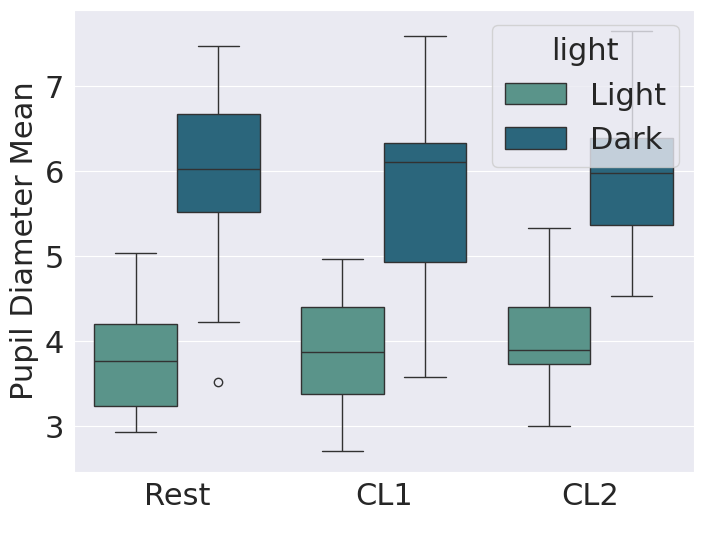} \\
(a) Pupil Diam. Mean & (b) Pupil Diam. Var & Pupil Diam. Mean \\
\includegraphics[width=0.33\linewidth, height=3.5cm]{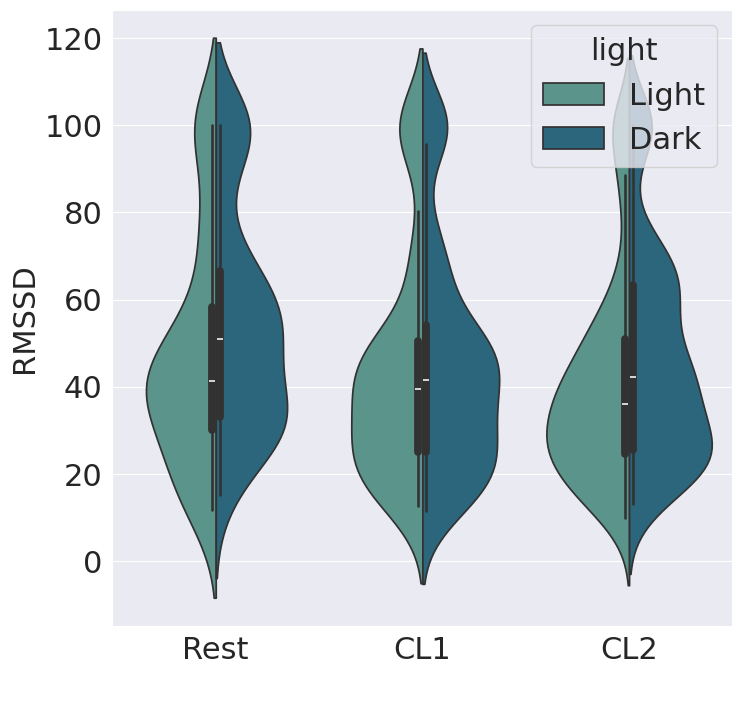} &
\includegraphics[width=0.32\linewidth, height=3.5cm]{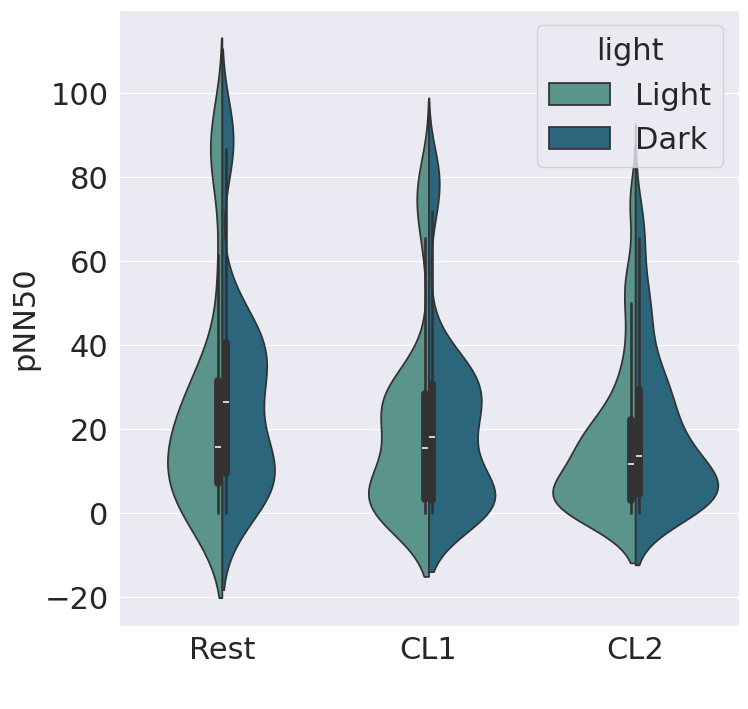} &
\includegraphics[width=0.32\linewidth, height=3.5cm]{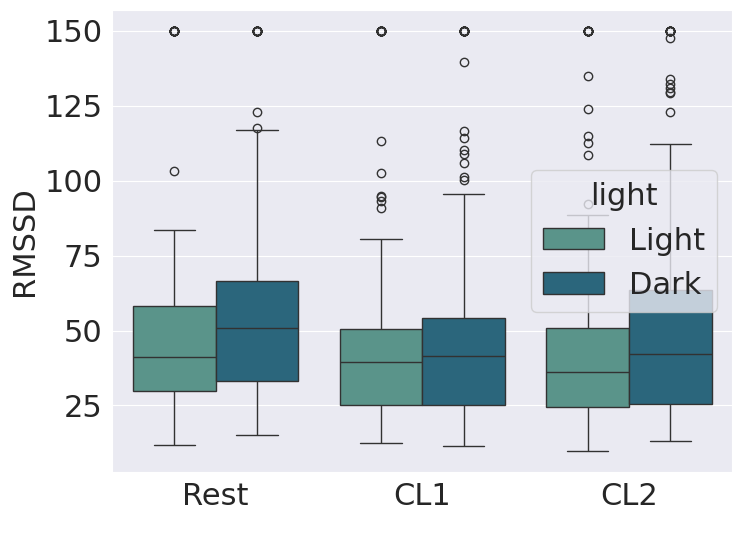} \\
(a) RMSSD & (b) pNN50 &  RMSSD \\
 \end{tabular}
 \caption{\textbf{Left}: Density plots of pupillometry and HRV features under light and dark conditions. The top row shows pupillometry features (mean pupil diameter, and pupil diameter variance), while the bottom row shows HRV features (RSMMD, and pNN50). Higher sensitivity to light of pupillometry features can be observed from their complete distributions' change. The HRV features are relatively less sensitive to light. \textbf{Right}: Box plot of the Pupil mean and RMSSD values for all cognitive load conditions using the Polar device. CL1 and CL2 stand for cognitive load for the 1-back (low) and 2-back (high) tasks respectively.}
 \label{fig:feat_violin_plot}
 \vspace{-0.6cm}
\end{figure*} 

To understand the feature level differences from varying light conditions, we selected the most common features from both pupillometry and HRV and compared their distribution using violin plots under light and dark conditions. Figure \ref{fig:feat_violin_plot} left shows the distribution under light and dark conditions. The pupillometry features exhibit a significant distribution shift between these conditions. In contrast, the distribution shift is significantly less for the HRV features, suggesting that combining both types of features can enhance robustness to light variations.

Figure \ref{fig:feat_violin_plot} right shows box plots of the most commonly used features mean pupil diameter and RMSSD. It reveals distinctly different distributions for light and dark conditions, with this difference being consistent across rest, 1-back, and 2-back conditions. This again confirms that pupil diameter is highly sensitive to lighting changes. For HRV, the difference between light and dark conditions is less pronounced, indicating that RMSSD is less sensitive to variations in lighting conditions.

\subsubsection{Statistical test results}
We conducted t-tests to understand  if the features analyzed above shows significant difference between light and dark conditions. The Mean Pupil Diameter shows a p-value of 7.164e-26, thus it has a significant difference between light and dark conditions. However, RMSSD obtains a p-value of 0.488, thus there is no statistically significant difference between light and dark conditions. More detailed t-tests comparing the effectiveness of Mean Pupil Diameter and RMSSD in distinguishing different cognitive load states are presented in table \ref{tab:ttest}. Here we observed that the Mean Pupil Diameter is unable to differentiate between CL1 and CL2 states whereas RMSSD is unable to differentiate between Rest and CL1 states. This again highlights the importance of having multimodal data to accurately differentiate the cognitive load levels.
\begin{table}
    \caption{Statistical test results comparing Pupil Diameter and RMSSD across all states (p-values are shown).}
    \centering\footnotesize
    \begin{tabular}{ l c c c}
    \hline
    \textbf{Feature} & \textbf{Rest vs CL1} & \textbf{Rest vs CL2} & \textbf{CL1 vs CL2} \\
    \hline
    \hline
    Pupil Diam & 0.0106 & 0.0052 & 0.1517 \\
    RMSSD & 0.3218 & 0.0010 & 0.0003\\
    \hline
    \end{tabular}
    \label{tab:ttest}
\end{table}

\subsection{Reliability over light conditions}
Another way to understand the challenge of brightness impact is to analyze the reliability of sensors in estimating cognitive load under varying light conditions. For this purpose, we computed the Expected Calibration Error (ECE) for individual sensors and the multimodal settings. ECE represents the expected absolute difference between the predicted confidence and the model accuracy \cite{ECE-Hauskrecht-2015}. Let $y$ be the true label, $\hat{y}$ be the predicted label, and $\hat{p}$ be the confidence of the predicted label. ECE is computed as $\textrm{E}_p\big[|\mathcal{P}(\hat{y}=y|\hat{p})-\hat{p}|\big]$. The results are shown in Table \ref{tab:ece_comparison}. It can be observed that HRV features report better ECE in estimating cognitive load under varying light conditions. The ECE of pupillometry-based cognitive load prediction is high, indicating unreliability under changing light conditions. In the multimodal settings, where HRV and pupillometry are used together, ECE shows significant improvement over the pupillometry-only setting.

\begin{table}
\caption{Expected Calibration Error(ECE) when using individual sensors and their combination (multimodal setting). The lower the ECE, the higher the reliability.}
\centering
\resizebox{.3\textwidth}{!}{%
\begin{tabular}{ c  c c c }
\hline
\textbf{Metric} & \textbf{HRV} & \textbf{Pupil} & \textbf{HRV+Pupil}\\  
\hline
\hline
\textbf{ECE} & 0.197 & 0.307 & 0.221 \\
\hline
\end{tabular}
}
\label{tab:ece_comparison}
\end{table}

\section{Discussion}
\label{sec:discussion}
Several aspects of the data processing pipeline are often assumed in the literature and in the available libraries for HRV and pupil data. Here, we discuss the impact of some key processing steps on classifier design and performance. To this end, we conducted a handful of ablation studies. Polar H10 is used as the HRV device for the studies discussed in this section. 

\subsection{Impact of window size for feature estimation}
The recommended window length for HRV features is 60 seconds \cite{60sec_window-Bernardes-2022}. When the window length is shorter, the computed features are typically less meaningful, especially the frequency domain. We compared the results from a 30-second window to a 60-second window, as shown in Table \ref{tab:30_60_sec_window}. The accuracy drops significantly for HRV features when the window size is 30 seconds (by 20.6 percentage points). As our feature importance analysis from the RF classifier highlights the significance of frequency-domain features, this drop in performance is expected without a precise estimation of those features. The pupillometry features show a smaller drop of 4.4\%, indicating they remain relatively reliable with a 30-second window. The overall performance drop in the multimodal settings is also more than 20\%. These results confirm the recommended window size of 60 seconds.
\begin{table}
    \caption{Performance comparison between 30 second and 60 second window size for feature estimation.}
    \centering\footnotesize
    \begin{tabular}{ l c c}
    \hline
    \textbf{Sensors} & \textbf{30 Sec} & \textbf{60 Sec} \\
    \hline
    \hline
    Pupil & 66.67$\pm$0.22 & 77.07$\pm$0.40 \\
    HRV & 69.05$\pm$0.24 & 82.15$\pm$0.28\\
    HRV+Pupil & 71.42$\pm$0.18 & 92.26$\pm$0.19\\
    \hline
    \end{tabular}
    \label{tab:30_60_sec_window}
\end{table}

\subsection{Impact of noise filtering on pupillometry data}
To cope with the noise in the pupil diameter data, we used filtering and smoothing with a butter low-pass filter of order 5 and cut-off 4Hz. Butter low-pass filters are commonly used in the pre-processing step for noise removal from pupillometry \cite{butter_filter-Knapen-2016}. Filtering with a low-pass filter helps us to remove high-frequency noise in the data. Smoothing helps to reduce the impact of short-term fluctuations. Table \ref{tab:filter_impact} shows the importance of filtering in our pre-processing pipeline. Without this filtering, the accuracy drops by more than 13.3\% points. Thus a significant fraction of the noise is filtered by the butter low-pass filter. 

\begin{table}
    \caption{Low-pass filter impact on RF classification accuracy on Experiment 1 (binary case) using all light conditions.}
    \centering\footnotesize
    \begin{tabular}{ l c}
    \hline
    \textbf{Settings} & \textbf{Accuracy} \\ 
    \hline
    \hline
    Without Butter Low-pass Filter & 57.81 \\ 
    With Butter Low-pass Filter & 77.05 \\ 
    \hline
    \end{tabular}
    \label{tab:filter_impact}
\end{table}

\section{Conclusion}
\label{sec:conclusion}
The aim of this research is to understand the influence of ambient light on pupillometry based CLE and how multimodal data acquisition helps to mitigate it. The multimodal in our case is realized using HRV alongwith pupillometry. We observed the robustness of pupillometry based CLE is low when light conditions are changed in the experimental settings. However, our study shows that by incorporating diverse lighting conditions in the training, the robustness to change in light conditions is improved. Combined with HRV, the multimodal data showed improved accuracy and robustness. HRV alone also has limitations due to its sensitivity to variations in the electromagnetic field and breathing cycle. The multimodal settings complements the limitations of any single modality, giving the best accuracy in all cases.

The HRV devices commonly used in cognitive load studies were chosen according to the experimental conditions. When controlled laboratory studies used high-fidelity HRV devices such as Biopac MP35, field application used fitness grade devices like Polar H10 or smartwatches. We compared both devices under the same task conditions by synchronously collecting data from Biopac MP35 and Polar H10. We then compared the CLE performance combining pupillometry with each device separately. The results show that using fitness grade Polar H10 is as good or sometimes better than clinical grade Biopac MP35, especially when the subjects can move during the data acquisition.

Our findings also reveal that neural network-based classifiers, particularly those using transformers, can achieve performance comparable to conventional random forest techniques. This is promising for future work, suggesting that advanced neural network architectures may offer robust solutions for cognitive load estimation, even in the presence of varying lighting conditions. The transformer-based classifier's ability to learn directly from raw signals without extensive feature engineering marks a significant advancement in the field.



\noindent \textbf{Acknowledgments}: This work was supported by the Natural Sciences and Engineering Research Council (NSERC) of Canada Discovery grant RGPIN-2020-06121.










\bibliographystyle{cas-model2-names}

\bibliography{cas-refs}



\end{document}